\documentclass[reprint,pra,aps,amsmath,amssymb]{revtex4-1}

\usepackage{graphicx}
\usepackage{dcolumn}
\usepackage{bm}
\usepackage{hyperref}
\hypersetup{colorlinks = true, linkcolor = [rgb]{0.19411,0.51882,0.667058}, urlcolor = [rgb]{0.125490,0.29542,0.1647058}, citecolor = [rgb]{0.75882,0.37411,0.14117}}

\graphicspath{{../images/}}

\usepackage[toc,page]{appendix}

\newcommand{\nn}{\nonumber \\}

\newcommand{\braket}[2]{\langle{#1}|{#2}\rangle}
\def\togli#1{}

\def\>{\rangle}
\def\<{\langle}

\usepackage{braket}

\def\>{\rangle}
\def\<{\langle}

\usepackage[bottom]{footmisc}

\usepackage{xcolor}

\begin{document}

\title{The quantum limit to incoherent imaging is achieved by linear interferometry}

\author{Cosmo Lupo}
\author{Zixin Huang}
\author{Pieter Kok}

\affiliation{Department of Physics \& Astronomy, University of Sheffield, Hicks Building, Hounsfield Road, Sheffield S3 7RH, United Kingdom}

\begin{abstract}
\noindent
We solve the general problem of determining, through imaging, the three-dimensional positions of $N$ weak incoherent point-like emitters in an arbitrary spatial configuration.
We show that a structured measurement strategy in which a linear interferometer feeds into an array of photo-detectors is always optimal for this estimation problem, in the sense that it saturates the quantum Cram\'er-Rao bound. We provide a method for the explicit construction of the optimal interferometer.
Further explicit results for the quantum Fisher information and the optimal interferometer design that attains it are obtained for the case of one and two incoherent emitters in the paraxial regime.
This work provides insights into the phenomenon of super-resolution through incoherent imaging that has attracted much attention recently.
Our results will find a wide range of applications over a broad spectrum of frequencies, from fluorescence microscopy to stellar interferometry.
\end{abstract}

\date{\today}
  
\maketitle

\noindent
Quantum imaging \cite{kolobov2007quantum} exploits quantum features of light to create an image of an object---or collection of objects---that emits or scatters light.
Advances in quantum imaging have followed several routes. Typically, the goal is to address the possibility of beating the limits of classical imaging \cite{born2013principles,rayleigh1879xxxi} by exploiting the unique properties of optical quantum states \cite{brida2010experimental,lugiato2002quantum,shapiro2012physics,Perez12,PhysRevA.77.043809,PhysRevA.79.013827, unternahrer2018super}.
For example, ghost imaging \cite{shapiro2012physics,erkmen2010ghost,PhysRevA.77.043809}, 
quantum lithography \cite{PhysRevLett.87.013602,PhysRevA.63.063407},
 and quantum sensing \cite{costa2006use,RevModPhys.89.035002,photonic_quantum_sensing} exploit entanglement to enable sensitivity and precision beyond what is achievable classically, whilst fluorescence super-resolution microscopy \cite{ram2006beyond, thorley2014super, small2014fluorophore,mortensen2010optimized} utilizes carefully engineered
 emitters and measurements to break the diffraction limit.

A renewed interest in the field was triggered recently by the work of Tsang, Nair, and Lu \cite{PhysRevX.6.031033}, who investigated the imaging of a pair of weak incoherent emitters in the far-field paraxial regime, such as a binary star system or a pair of fluorescent emitters. 
They considered the problem of measuring, through imaging, the transverse angular separation between the two sources,
and used the tools of quantum estimation theory, in particular, the quantum Fisher information (QFI) and the Cram\'er-Rao bound \cite{paris2009quantum, Sidhu}.
They showed that a structured measurement setup in which the light focused on the image plane is first passed through a Hermite-Gaussian mode sorter is superior to direct imaging.
The statistical error for the estimation of the transverse angular separation between two identical sources is constant (independent of the separation), and inversely proportional to the Rayleigh length.

By contrast, direct detection sees the error in the estimation of the angular separation increase substantially when the separation between the sources falls below the Rayleigh length, a phenomenon dubbed the ``Rayleigh curse''.
The method for obtaining sub-Rayleigh super-resolution through coherent detection of incoherent light has been further developed, generalized \cite{PhysRevLett.117.190802,PhysRevLett.117.190801,PhysRevA.95.063829,PhysRevA.96.063829,PhysRevA.95.063847,PhysRevA.96.062107,PhysRevLett.121.023904,PhysRevLett.117.190801,PhysRevLett.121.180504,PhysRevLett.122.140505,npjQI2019, PhysRevA.99.033847,PhysRevA.99.012305}, and demonstrated \cite{paur2016achieving,tang2016fault,yang2016far,PhysRevLett.118.070801,PhysRevLett.121.090501,PhysRevLett.121.250503,paur2018tempering,hassett2018sub,zhou2019quantum}.
See Ref.~\cite{tsang2019resolving} for a comprehensive review on recent progress and related topics.

The problems considered so far have been limited, with a few exceptions \cite{tsang2017subdiffraction,PhysRevA.99.013808,PhysRevA.99.033847,ADatta}, to a pair of point-like emitters. There is still no general quantum theory that can be applied to a situation where an arbitrary number of emitters lay within a region of the size of the Rayleigh length.
Furthermore, we still lack a general insight into why interferometric measurements are optimal for this family of estimation problems. 
In this paper, we answer both these questions: (i) we determine the QFI for the three-dimensional positions of any number of point sources in an arbitrary spatial configuration; (ii) we show that a structured measurement strategy where a linear interferometer feeds into an array of photo-detectors is always optimal for this general estimation problem. We provide an explicit construction for the interferometer, which then can be realized using standard techniques \cite{Reck,Clements}.

Our theory is based on a very general model for the optical system that is used to collect and measure light. It includes as special cases the standard imaging model based on the point-spread function, as well as interferometric measurements as stellar interferometry \cite{monnier2003optical,lawson2000principles}.

\begin{figure}[t]
\includegraphics[width=8cm]{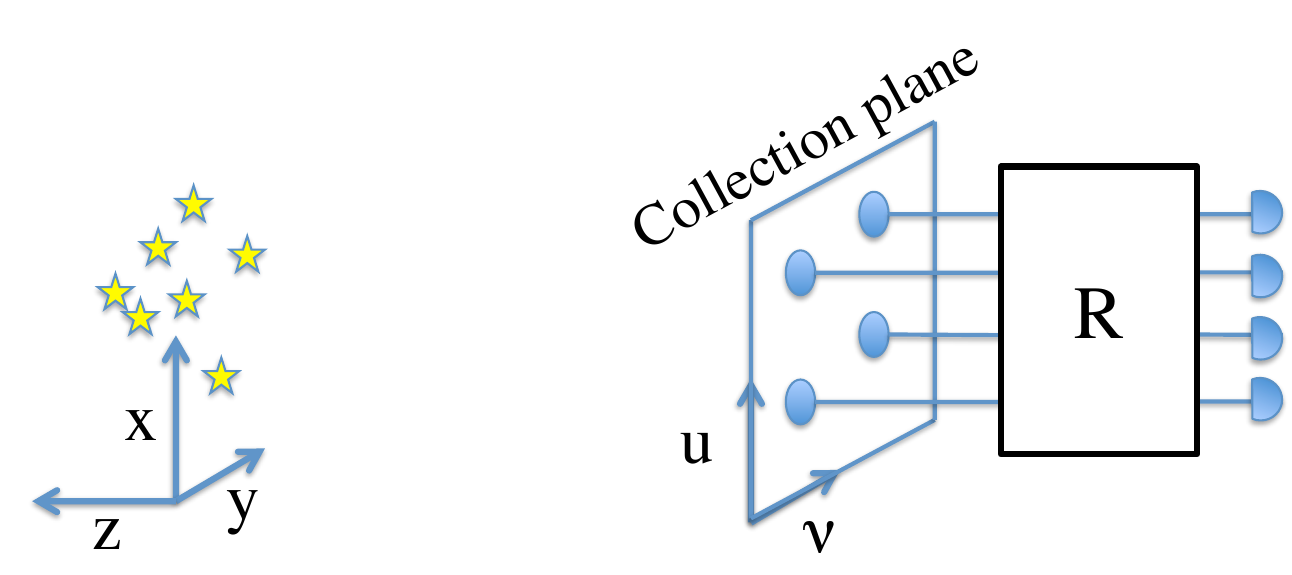} 
\caption{A collection of $N_S$ incoherent point-like emitters (left) and the apparatus used to measure them (right). The light emitted or scattered by the objects is collected at $N_C$ specific locations in the collection plane. The collected light is coherently processed in a general interferometer and measured using photo-detection.}\label{scheme_fin}
\end{figure}

{\it The Model:}
Consider a system of $N_S$ point-like objects that emit or scatter quasi-monochromatic, incoherent light. The system is measured by collecting the light that impinges on a system of $N_C$ \emph{collectors} with accurately known positions. Schematically, these can be considered as pinholes on a light-collection plane (see Fig.~\ref{scheme_fin}), microlenses coupled into optical fibers, the input of a photonic lantern \cite{phlantern}, or telescope arrays such as the Very Large Telescope.
The collected light is coherently processed in a general device $R$, and measured using photodetectors.
If the collectors are arranged to form a square array, they can be used to model the pixel of a CCD camera. 
Taking a continuous distribution of collectors, one can recover the thin lens model of an optical imaging system.

A single source $s$ has coordinates $r_s = (x_s,y_s, z_0 + z_s)$, where the first two are the transverse coordinates and the third component lies along the optical axis. 
Here $z_0$ is a reference distance between the objects and the collection plane.
For simplicity, we assume that the collectors lie in a transverse plane (this constraint may be relaxed), with collector $j$ having coordinates $w_j = (u_j,v_j,0)$. We work in the limit of weak sources and assume that at most one photon is collected in one detection window.

The state of a photon emitted by source $s$ impinging on the $N_C$ collectors is described by 
\begin{align}\label{1photonwf}
|\psi(r_s) \rangle = \sum_{j=1}^{N_C} \gamma(w_j,r_s) |j \rangle \, ,
\end{align}
where $|j\rangle$ denotes the state of a photon arriving at collector $j$, and $\gamma(w_j,r_s)$ is the corresponding complex amplitude.
This model requires that the light coupled to a collector is described as a single mode.
In general, the phase of $\gamma(w_j,r_s)$ is expressed by the optical path length from the source to the collector:
\begin{align}
\arg \gamma  
= k \sqrt{ \left( x_s - u_j \right)^2 + \left( y_s - v_j \right)^2 + \left( z_0 + z_s \right)^2 } \, ,
\end{align}
where $k$ is the wave number. The modulus of $\gamma$ is inversely proportional to the distance between the source and the collector. The normalization condition is $\sum_{j} |\gamma(w_j,r_s)|^2 = 1$.
The total state of a single photon coming from $N_S$ weak incoherent sources 
is given by 
\begin{align}\label{mixedstate}
\rho(r) = \sum_{s=1}^{N_S} p(s) |\psi (r_s) \rangle \langle \psi (r_s) | \, ,
\end{align}
where $r \equiv (x_1,y_1,z_1\dots,x_{N_S},y_{N_S},z_{N_S})$ indicates the collective coordinates of the $N_S$ emitters, and $p(s)$ is the probability that the photon is emitted by source $s$.


{\it Quantum Fisher Information:}
We are interested in measuring one or more generalized coordinates of the system of $N_S$ emitters.
Consider a unit vector with $3N_S$ components
\mbox{$a = (a_1, a_2, \dots, a_{3N_S})$}.
A generalized coordinate is defined as $\vartheta := a \cdot r$, i.e., the scalar product of a unit vector with the collective coordinates of the object.
For a variation $\delta\vartheta$ in the parameter $\vartheta$, the collective coordinate changes from $r$ to $r' = r + a \delta\vartheta$, with $\delta\vartheta = a \cdot (r' - r)$. We first obtain a simple expression for the QFI of $\vartheta$ in our model. Second, we compute the classical Fisher information (CFI) for the case where $R$ is a linear interferometer followed by photo-detectors. Third, we construct an $R$ such that the CFI equals the QFI, proving that linear interferometry and photo-detection are optimal for the measurement of $\vartheta$.

To compute the QFI, consider a purification of the mixed state $\rho(r)$ in Eq.~(\ref{mixedstate}). We introduce an auxiliary $N_S$-dimensional Hilbert space spanned by a set of orthogonal unit vectors $ |s\rangle $ such that
\begin{align}\label{pur}
|\Psi(r) \rangle & = \sum_{j,s} c(w_j,r_s) |j \rangle |s\rangle \, ,\\
c(w_j,r_s) &:= \sqrt{p(s)} \,  \gamma(w_j,r_s) \, .
\end{align}
%
The QFI $I_Q(\vartheta)$ is given by the relation
\begin{equation}\label{Uhlmann}
I_Q(\vartheta) = \lim_{\delta\vartheta \to 0} \frac{ 8( 1 - f_{r,r'} ) }{ \delta\vartheta^2 } \, ,
\end{equation}
where
\begin{equation}
\label{eq:fid}
f_{r,r'} = \max_{V} |\langle \Psi(r) | (I \otimes V) | \Psi(r') \rangle |\, ,
\end{equation}
is the Uhlmann fidelity \cite{NielsenChuang}, and the maximization is over all unitary transformations $V$ acting on the purifying system.
By substituting the expression for the purification in Eq.~(\ref{pur}) into Eq.~\eqref{eq:fid}
we obtain an explicit expression for the fidelity:
\begin{align}
f_{r,r'} & = \max_V  \left| \sum_{st} V_{st} \sum_j c(w_j,r_s)^* c(w_j,r'_t) \right| \\
& = \max_V \left| \mathrm{Tr}( V^\mathsf{T} M) \right| 
= \| M \|_1 \, , \label{qfid}
\end{align}
where the last equality follows from a variational expression of the trace norm \cite{NielsenChuang}, $\| M \|_1 = \mathrm{Tr}( \sqrt{M^\dag M} )$, and we have defined the matrices
\begin{align}
V_{st}  := \langle s | V | t \rangle \, , \quad
M_{st} : = \sum_u c(w_j,r_s)^* c(w_j,r'_t) \, .
\end{align}
Therefore, we have shown that the QFI can be expressed as a function of the matrix $M$ (similar to Ref.~\cite{Quantum_metrology_matrix}) that characterizes the geometry of the light sources and the collectors.

Next, we obtain an expression for the CFI for $\vartheta$ from a measurement comprising a linear interferometer characterized by a $N_C \times N_C$ unitary matrix $R$ and an array of photo-detectors (shown in Fig.~\ref{scheme_fin}).
Given the state $\rho(r)$ in Eq.~(\ref{mixedstate}), the probability of observing a photon at detector $q$ is
\begin{align}
p_q = \sum_s    \left| \sum_j    c(w_j,r_s) R_{jq} \right|^2 \, .
\end{align}
Given our assumption that no more than one photon arrives at the collector plane within one detection time period, at most one detector will click.
The classical fidelity between two probability distributions corresponding to source configurations $r$ and $r'$ is
\begin{align}\label{cfid}
f^c_{r,r'} = \sum_q \sqrt{ \sum_{s,t}    \left| \sum_j    c(w_j,r_s) R_{jq} \right|^2 
\left| \sum_{j'} c(w_{j'},r'_{t}) R_{j'q} \right|^2 } \, .
\end{align}
The CFI for $\vartheta$ is then given by the classical version of Eq.~(\ref{Uhlmann}). To find the optimal interferometer, we minimize the classical fidelity in Eq.~(\ref{cfid}) over the set of unitary matrices $R$. The classical fidelity is lower bounded by the quantum fidelity in Eq.~(\ref{qfid}),
$f^c_{r,r'} \geq f_{r,r'}$,
which is an instance of the Cauchy-Schwarz inequality \cite{fuchs1995mathematical}. The optimality of the measurement based on interferometry and photo-detection then follows from the fact that there exists a choice of $R$ that saturates the Cauchy-Schwarz inequality. We show this explicitly in the Supplementary Material. Our proof gives an explicit construction of the interferometer matrix $R$. 
This result has been obtained without invoking the paraxial approximation.

Note that, in general, the optimal $R$ may depend on the value of the parameter. This is a feature that is often encountered in quantum metrology and implies that the optimal measurement cannot be decided from the outset \cite{Sidhu}.
However, in the paraxial regime, the optimal interferometer does not depend on the parameter for the problem of estimating the separation between two sources.

{\it Paraxial Regime:}
Of particular interest to astronomy is the paraxial regime, where $x_s, y_s \ll z_0$. We also assume that $z_s \ll z_0$.
In this approximation the optical path length from a source at location $r_s$ to collector at $w_j$ becomes
\begin{align}
 \arg\gamma( w_j , r_s) & \simeq -  k\frac{ u_j x_s + v_j y_s }{z_0} -  k \frac{z_s}{z_0} \frac{ u_j^2 + v_j^2 }{2z_0} \cr 
  & =:  \phi( w_j , r_s)  \, ,
\end{align}
where we have kept only the terms that are linear in $r_s/z_0$. We neglected the global phases that depend only on $r_s$ since the sources are incoherent, and the terms that depend only on $w_j$ are absorbed in the definition of the single photon states $|j\rangle$. 
In this regime, we may assume that all sources have the same distance from the collection plane, such that
$\gamma( w_j, r_s) = N_C^{-1/2} e^{i\phi( w_j , r_s) }$.
We introduce the operators:
\begin{align}
\hat g_x = \frac{k \hat u}{z_0}\, , \quad
\hat g_y = \frac{k \hat v}{z_0} \, , \quad
\hat g_z = \frac{1}{2} \frac{k (\hat u^2 + \hat v^2) }{z_0^2} \, ,
\end{align}
where $\hat u$, $\hat v$ are position operators of the collectors.
%
These operators are the generators of an Abelian unitary group. For each source coordinate $r_s = (x_s, y_s, z_s)$ we define the unitary operator
\begin{align}
U(r_s) = e^{- i \hat g_x x_s - i \hat g_y y_s - i \hat g_z z_s } \, .
\end{align}
In the paraxial regime this unitary generates a specific instance of the single-photon wave function of Eq.~(\ref{1photonwf}):
\begin{align}
|\psi(r_s) \rangle  &= U(r_s) |\psi(0) \rangle \nn
& = \frac{1}{\sqrt{N_C}} \sum_{j} e^{i \phi(w_j,r_s)} |j \rangle \cr
& = \frac{1}{\sqrt{N_C}} \sum_{j} e^{ - i k\frac{ u_j x_s + v_j y_s }{z_0} - i k \frac{z_s}{z_0} \frac{ u_j^2 + v_j^2 }{2z_0} } |j \rangle 
\label{dummy}
\end{align}
with $|\psi(0) \rangle : = N_C^{-1/2} \sum_{j} |j \rangle$ a reference state that does not depend on $r_s$.
This representation is readily extended to $N_S$ incoherent sources via Eq.~(\ref{mixedstate}):
\begin{align}
\rho(r) 
& =  \sum_s p(s)|\psi (r_s) \rangle \langle \psi (r_s) | \cr
& = \sum_s p(s) U(r_s) |\psi(0)  \rangle \langle \psi(0)| U(r_s)^\dag  \, .
\end{align}
It allows us to compute directly the QFI matrix for the source coordinates.

As an example, for two identical emitters, we estimate the centroid and the relative coordinates. 
We find that their QFI matrix is proportional to the covariance matrix of the generators (see Supplementary Material):
\begin{align}
\sigma_{ab} = \langle \hat g_a \hat g_b \rangle - \langle \hat g_a \rangle \langle \hat g_b \rangle \, ,
\end{align}
for $a,b = x,y,z$, where the average is over the dummy state $|\psi(0)\rangle$ introduced in Eq.~(\ref{dummy}). This covariance matrix is a function of the spatial distribution of the collectors only. We can compute it explicitly, for example, for a continuous distribution of collectors that define a circular aperture, and we reproduce the results of Yu and Prasad \cite{PhysRevLett.121.180504}.

For the estimation of a transverse separation $\Delta x$ along one transverse direction, we obtain the following expression for the QFI:
\begin{align}
I_Q(\Delta x)
& =  \langle \hat g_x^2 \rangle - \langle \hat g_x \rangle^2  =  \frac{k^2}{z_0^2} \left( \langle \hat u^2 \rangle - \langle \hat u \rangle^2 \right) .
\end{align}
This shows that the accuracy of the estimation is characterized by the variance of the spatial distribution of the collectors.
For a continuous distribution of collectors that simulate the aperture of a microscope or telescope, the variance is proportional to the square of the aperture size $D$, which yields $I_Q(\Delta x) \sim \frac{k^2 D^2}{z_0^2} \sim \mathrm{x_R}^{-2}$, where $\mathrm{x_R}$ is the Rayleigh length of the optical imaging system. 


\begin{figure}[t]
\includegraphics[trim = 0cm 0cm 0cm 0cm, clip, width=1.0\linewidth]{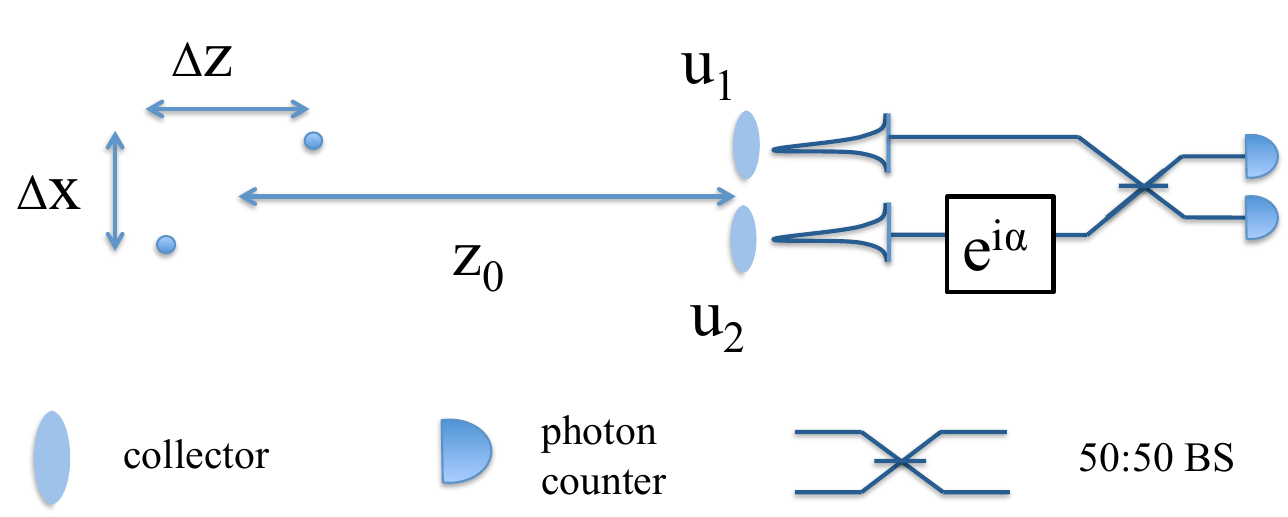} 
  \caption{Schematic of two sources with a separation of $\Delta x$ in the object plane, and a separation $\Delta z$ in the axial direction. Two collectors at $u_1$ and $u_2$ direct light into a two-mode interferometer consisting of a phase shift $\alpha$ and a 50:50 beam splitter, followed by two photon counters. }
  \label{f:phase_est}
\end{figure}

{\it Physical Implementation:}
Consider the simplest case of using two collectors to estimate the transverse angular separation $\Delta x$ of two sources. The collectors are placed at positions $u_1$ and $u_2$, and the precision in $\Delta x$ is given by (see Supplemental Material):
\begin{align}
I_Q(\Delta x) = \frac{k^2}{4 z_0^2} \left( u_1 - u_2\right)^2.
\end{align}
A simple experimental scheme that can achieve this precision is given in Fig.~\ref{f:phase_est}, for a phase shift $\alpha = 0$. It is worth noting that our expressions for the QFI and CFI are for estimating the linear separation $\Delta x$, instead of the angular separation, which would be 
${\Delta x}/{z_0}$. If we were to estimate the angular separation, the expression for the QFI would increase by a factor of $z_0^2$, i.e $ I_Q(\Delta x/z)\approx {(u_1-u_2)^2}/{4}$. 
We thus reproduce the celebrated result of Ref.~\cite{PhysRevX.6.031033}, using a simple scheme interferometry. Such a scheme is also optimal for other imaging applications \cite{Pearce2017optimalquantum,Howard19}. 
Note that the optimal interferometer does not depend on the parameter for this example.


One may wish to also estimate the axial separation $\Delta z$ simultaneously with $\Delta x$, as considered in Refs.~\cite{PhysRevLett.121.180504,PhysRevLett.122.140505}. However, using only two collectors, one cannot extract two parameters from one measured degree of freedom.
A necessary condition to simultaneously estimate the parameters $\Delta x_a$ and $\Delta x_b$, is that the QFI sub-matrix is diagonal, i.e., we must use a spatial distribution of the collectors such that, for $a \neq b$,
\begin{align}
\sigma_{ab}=\braket{\hat g_a \hat g_b}  - \braket{\hat g_a} \braket{\hat g_b} = 0\, .
\label{eq:off_diag}
\end{align}
One such configuration is four evenly spaced collectors on a line. We find that the optimal interferometer is simply a quantum Fourier transform of the four modes (Fig.~\ref{f:4_det}), independent of the parameters.

{\it Discussion and conclusions:}
Recent developments have shown that coherent detection schemes are optimal for estimating the transverse separation between two point-like weak incoherent sources.
In particular, these schemes sidestep the so-called ``Rayleigh curse'' that limits the precision of direct detection in the sub-Rayleigh regime. 

\begin{figure}[t] \center
\includegraphics[trim = 0cm 0cm 0cm 0cm, clip, width=1.0\linewidth]{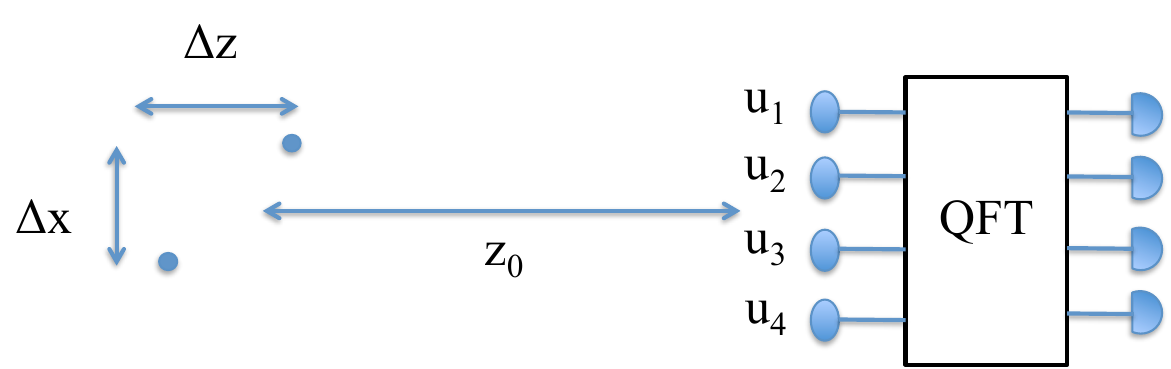} 
  \caption{\label{f:4_det} Schematic for estimating simultaneously $\Delta x$ and $\Delta z$ using four collectors which are evenly spaced, centered at position 0. The optimal linear optical transformation is a four-mode quantum Fourier transform.}
\end{figure}

Here, we solve the general problem of determining the three-dimensional positions of $N_S$ weak incoherent point-like emitters in an arbitrary spatial configuration.
We introduce a general model where light is collected at different locations, and is coherently measured.
Our model includes stellar interferometry and imaging through a circular aperture in the limit of a continuous distribution of collectors.

Our analysis shows that linear interferometry and photon counting are always optimal for estimating generalized coordinates of the sources.
Furthermore, we provide an explicit construction for the optimal interferometer, which then can be implemented using standard methods \cite{Reck,Clements}. 

Our results explain why coherent detection overcomes the Rayleigh curse by recasting imaging as interferometry at the outset. We have shown that, for the case of two incoherent sources, the optimal interferometer can have relatively low complexity, for example, a single beam-splitter, and a quantum Fourier transform for the case of four collectors. 
In general, the optimal interferometer may depend on the parameter to be estimated, but our examples show that it is parameter-independent for the problem of estimating the separation between two point sources.

A number of questions remain open. For example: under what conditions are the optimal interferometer independent of the parameter to be estimated? When is it possible to use the same interferometer for the optimal estimation of multiple parameters?
Finally, we expect that our approach can be generalized and applied to extended sources \cite{PhysRevA.99.033847}, non-weak or non-incoherent sources \cite{PhysRevLett.117.190802}, and to estimate other parameters beyond generalised spatial coordinates, for example spatial moments \cite{tsang2017subdiffraction}, as well as time and frequency measurements \cite{SilberhornPRL2018,AlexRetzker}.

{\it Acknowledgements:}
This work was supported by the EPSRC Quantum Communications Hub, Grant  No.EP/M013472/1. The authors thank Mankei Tsang, Tommaso Tufarelli and Mark Wilde for useful comments on the manuscript.

\newpage

\appendix 

\section{Optimality of linear optics and photo-detection}

In the main body of the paper, we have obtained expressions for the quantum Fisher information (QFI) and for the classical Fisher information (CFI) for an estimation strategy where the light at the collectors is first processed coherently by a linear interferometer and then measured by photo-detection. Here we show that, for a suitable choice of the interferometer matrix, the CFI equals the QFI, therefore showing the optimality of said measurement.

Before proceeding to the proof, we recall that the QFI is given by the following expression:
\begin{equation}
I_Q(\vartheta) = \lim_{\delta\vartheta \to 0} \frac{ 8( 1 - f_{r,r'} ) }{ \delta\vartheta^2 } \, ,
\end{equation}
where the Uhlmann fidelity
\begin{align}
f_{r,r'} = \| M \|_1 \, , 
\end{align}
is given by the trace norm of the matrix $M$ with components
\begin{align}
M_{st} = \sum_j c(w_j,r_s)^* c(w_j,r'_t) \, .
\end{align}

On the other hand, for a given interferometer characterized by a $N_C \times N_C$ unitary matrix $R$, the CFI reads
\begin{equation}
I(\vartheta) = \lim_{\delta\vartheta \to 0} \frac{ 8( 1 - f_{r,r'} ) }{ \delta\vartheta^2 } \, ,
\end{equation}
where the classical fidelity is given by the following expression: 
\begin{align}\label{cfid}
f^c_{r,r'} = \sum_q \sqrt{ \sum_{s,s'}    \left| \sum_j    c(w_j,r_s) R_{jq} \right|^2 
\left| \sum_{j'} c(w_{j'},r'_{s'}) R_{j'q} \right|^2 } \, .
\end{align}

From the very definition of the QFI it follows that
\begin{align}\label{CS0}
f^c_{r,r'} \geq f_{r,r'} \, .
\end{align}

We are now ready to present the optimality proof, which is divided into two parts. First, we show that inequality (\ref{CS0}) is a Cauchy--Schwarz inequality. Second, we show that there exists a choice of $R$ that saturates it.

\subsection{Cauchy--Schwarz inequality}

The matrix $M$ can be written as follows:
\begin{align}
M = C(r)^\dag C(r') \, ,
\end{align}
where
\begin{align}
C(r) := \left(
\begin{array}{cccc}
c(w_1,r_1)   & c(w_1,r_2)   & \dots & c(w_1,r_{N_S}) \\
c(w_2,r_1)   & c(w_2,r_2)   & \dots & c(w_2,r_{N_S}) \\
\dots      & \dots      & \dots & \dots \\
c(w_{N_C},r_1) & c(w_{N_C},r_2) & \dots & c(w_{N_C},r_{N_S}) 
\end{array}
\right) \, ,
\end{align}
and similarly 
\begin{align}
C(r') := \left(
\begin{array}{cccc}
c(w_1,r'_1)   & c(w_1,r'_2)   & \dots & c(w_1,r'_{N_S}) \\
c(w_2,r'_1)   & c(w_2,r'_2)   & \dots & c(w_2,r'_{N_S}) \\
\dots      & \dots      & \dots & \dots \\
c(w_{N_C},r'_1) & c(w_{N_C},r'_2) & \dots & c(w_{N_C},r'_{N_S}) 
\end{array}
\right) \, ,
\end{align}

The trace norm $\| M \|_1$ can be obtained from the singular value decomposition of $M$, i.e.\ by finding the unitary transformations $V$ and $W$ that make $M$ diagonal:
\begin{align}\label{SVD}
V^\dag M W = D \, , 
\end{align}
where $D$ is a diagonal, non-negative matrix, yielding $\| M \|_1 = \mathrm{Tr}D$.

We then have, for any given unitary matrix $R$,
\begin{align}
\| M \|_1 
& = \mathrm{Tr}D \\
& = \mathrm{Tr}(V^\dag M W) \\
& = \mathrm{Tr}(V^\dag C(r)^\dag C(r') W) \\
& = \mathrm{Tr}(V^\dag C(r)^\dag R R^\dag C(r') W) \\
& = \sum_{s,s',j,j',t,v} V^*_{st} \, c(w_j,r_s)^* R_{vj}^* R_{vj'} c(w_{j'},r'_{s'}) W_{s't}
    \, .
\end{align}

\begin{widetext}

By applying the Cauchy--Schwarz inequality $N_C$ times we obtain
\begin{align}
\| M \|_1 & = \sum_v \left\{ \sum_t \left[ \sum_{j,s} R_{vj} c(w_j,r_s) V_{st}  \right]^* 
                                    \left[ \sum_{j',s'} R_{vj'} c(w_{j'},r'_{s'}) W_{s't} \right] \right\} \\
& \leq \sum_v \sqrt{ \sum_t \left| \sum_{j,s}  R_{vj}  c(w_j,r_s)V_{st}  \right|^2 }
              \sqrt{ \sum_{t'} \left| \sum_{j',s'} R_{vj'} c(w_{j'},r'_{s'}) W_{s't'}    \right|^2 } \, .
\end{align}

The last step is to note that the quantity on the right-hand side is invariant under the unitary transformations $V$ and $W$. 
Therefore, putting $V = W = I$ we finally obtain inequality (\ref{CS0}): 
\begin{align}
f_{r,r'} = \| M \|_1 
& \leq \sum_v \sqrt{ \sum_t \left| \sum_{j} R_{vj}  c(w_j,r_t)  \right|^2 }
\sqrt{ \sum_{t'} \left| \sum_{j'} R_{vj'} c(w_{j'},r'_{t'})    \right|^2 }
%
= f^c_{r,r'} \, .
\end{align}

In conclusion, we have shown that inequality (\ref{CS0}) is nothing but an instance of the Cauchy--Schwarz inequality.

\subsection{Saturation of the Cauchy--Schwarz inequality}

To conclude the optimality proof we will now show that there exists a choice of $R$ that saturates inequality (\ref{CS0}).

Consider the matrix
\begin{align}
A(r) := C(r) V = \left(\begin{array}{cccc}
a(w_1,1) & a(w_1,2) & \dots & a(w_1,{N_S}) \\
a(w_2,1) & a(w_2,2) & \dots & a(w_2,{N_S}) \\
\dots & \dots & \dots & \dots \\
a(w_{N_C},1) & a(w_{N_C},2) & \dots & a(w_{N_C},{N_S}) 
\end{array}\right) \, ,
\end{align}
where $V$, as well as $W$ below, is the unitary matrix that appears in the singular value decomposition of $M$ in Eq.\ (\ref{SVD}).

This matrix can be seen as a list of $N_S$ vectors $a(1),a(2), \dots, a(N_S)$, where
\begin{align}
a(s) = \left(\begin{array}{c}
a(w_1,s) \\
a(w_2,s) \\
\dots \\
a(w_{N_C},s) 
\end{array}\right) \, .
\end{align}

Similarly we define the matrix
\begin{align}
B(r') := C(r') W 
= \left(\begin{array}{cccc}
b(w_1,1) & b(w_1,2) & \dots & b(w_1,{N_S}) \\
b(w_2,1) & b(w_2,2) & \dots & b(w_2,{N_S}) \\
\dots & \dots & \dots & \dots \\
b(w_{N_C},1) & b(w_{N_C},2) & \dots & b(w_{N_C},{N_S}) 
\end{array}\right) \, ,
\end{align}
and the vectors $b(1),b(2), \dots, b(N_S)$, where
\begin{align}
b(s) = \left(\begin{array}{c}
b(w_1,s) \\
b(w_2,s) \\
\dots \\
b(w_{N_C},s) 
\end{array}\right) \, .
\end{align}

Note that the condition $V^\dag C(r)^\dag C(r') W = D$ determines the 
scalar products of the $a$'s vectors with the $b$'s vectors, i.e.,
\begin{align}
\sum_{j=1}^{N_C} a(w_j,s)^* b(w_j,t) = D_s \delta_{st} \, .
\end{align}

The matrix $R$ transforms the matrix $A$ into $A'$:
\begin{align}
A'(r) = R A(r) \, .
\end{align}
It follows from well-known results in linear algebra (QR decomposition) that there exists a choice of $R$ that transforms $A(r)$ into an upper-triangular matrix, i.e.,
\begin{align}
A'(r) = R A(r) =
\left(\begin{array}{ccccc}
a'(1,1) & a'(1,2) & a'(1,3) & \dots & a'(1,N_S) \\
0       & a'(2,2) & a'(2,3) & \dots & a'(2,N_S) \\
0       & 0       & a'(3,3) & \dots & a'(3,N_S) \\
0       & 0       & 0       & \dots & a'(4,N_S) \\
\dots   & \dots   & \dots   & \dots & \dots     \\
0       & 0        & \dots   & \dots & a'(N_C,N_S) \\
\dots   & \dots   & \dots   & \dots & \vdots       \\
0       & 0       & 0       & \dots & 0
\end{array}\right)
\, .
\end{align}
Note that here we are implicitly using the assumption that the number of collectors $N_C$ is at least equal to the number of sources $N_S$.

Because $R$ preserves the scalar product, the matrix $B$ is necessarily transformed into a lower-triangular matrix, i.e.,
\begin{align}
B'(y) = R B(y) =
\left(\begin{array}{ccccc}
b'(1,1)   & 0         & 0         & \dots & 0  \\
b'(1,2)   & b'(2,2)   & 0         & \dots & 0         \\
b'(1,3)   & b'(2,3)   & b'(3,3)   & \dots & 0         \\
\dots     & \dots     & \dots     & \dots & \dots     \\
b'(1,N_S) & b'(2,N_S) & b'(3,N_S) & \dots & b'(N_S,N_S) \\
\dots     & \dots     & \dots     & \dots & \dots     \\
b'(1,N_C) & b'(2,N_C) & b'(3,N_C) & \dots & b'(N_S,N_C)
\end{array}\right)
\, .
\end{align}

The scalar product then reads:
\begin{align}
D_s = \sum_{j=1}^{N_C} a(j,s)^* b(j,s) = \sum_{j=1}^{N_C} a'(j,s)^* b'(j,s) = a'(s,s)^* b'(s,s)\, .
\end{align}
As the coefficients $D_s$ are real and non-negative, we also have 
\begin{align}
D_s = \left| a'(s,s)^* b'(s,s) \right| = \left| a'(s,s) \right| \left| b'(s,s) \right| \, .
\end{align}

We can then write the quantum fidelity as follows:
\begin{align}\label{fdiag}
f_{r,r'} = \| M \|_1 = \sum_s D_s =  \sum_s \left| a'(s,s) \right| \left| b'(s,s) \right| \, .
\end{align}
On the other hand, we can write, for the same choice for the matrix $R$, the classical fidelity:
\begin{align}
f^c_{r,r'} & = \sum_v \sqrt{ \sum_t \left| \sum_{j,s}  R_{vj}  c(w_j,r_s)V_{st}  \right|^2 }
              \sqrt{ \sum_{t'} \left| \sum_{j',s'} R_{vj'} c(w_{j'},r_{s'}) W_{s't'}    \right|^2 } \\
& =  \sum_v \sqrt{ \sum_t \left| a'(v,t) \right|^2 }
              \sqrt{ \sum_{t'} \left| b'(v,t') \right|^2 } \\
& =  \sum_v \left| a'(v,v) \right| \left| b'(v,v) \right| \, . \label{fcdiag}
\end{align}
In conclusion, a direct comparison of Eq.\ (\ref{fdiag}) and Eq.\ (\ref{fcdiag}) verifies that $f_{r,r'} = f_{r,r'}^c$ for the above choice of $R$.

Note that the proof also gives an explicit construction for the matrix $R$.

\section{Paraxial regime}

In this section, we solve explicitly examples of estimation of one emitter and two incoherent emitters.

We apply the theory developed in the main body of the paper, where we have shown that the wave function of a single emitter at location $r = (x,y,z)$ can be written as 
\begin{align}\label{onesource}
|\psi(r) \rangle 
= U(r) |\psi(0) \rangle \, ,
\end{align}
where
\begin{align}
U(r) = e^{- i \hat g_x x - i \hat g_y y - i \hat g_z z } \, ,
\end{align}
and 
\begin{align}
|\psi(0) \rangle 
= \frac{1}{\sqrt{N_C}} \sum_{j=1}^{N_C} |j \rangle \, .
\end{align}

The unitary $U(r)$ has generators 
\begin{align}
\hat g_x = \frac{k \hat x}{z_0} \, , \, 
\hat g_y = \frac{k \hat y}{z_0} \, , \,
\hat g_z = \frac{1}{2}\frac{k (\hat x^2 + \hat y^2) }{z_0^2} \, .
\label{eq:gens}
\end{align}

Similarly, the state of $N_S$ incoherent source reads
\begin{align}
\rho(r) 
= \frac{1}{N_S} \sum_s |\psi (r_s) \rangle \langle \psi (r_s) | 
= \frac{1}{N_S} \sum_s U(r_s) |\psi(0)  \rangle \langle \psi(0)| U(r_s)^\dag  \, .
\end{align}

\subsection{One source}

Consider first the case of a single point-like source. Our goal is to compute the QFI matrix for its three spatial coordinates. 

According to Eq.\ (\ref{onesource}), the single-photon state depends on the source coordinates through a unitary transformation. Therefore, general results on quantum estimation theory yield that the QFI matrix is four times the covariance matrix of the generators. Therefore, the elements of the QFI matrix are:
\begin{align}
\mathrm{QFI}_{xx} & = 4 \left( \langle \hat g_x^2 \rangle - \langle \hat g_x \rangle^2 \right) 
          = \frac{4k^2}{z_0^2} \left( \langle u^2 \rangle - \langle u \rangle^2 \right) \, , \label{QFIm1}\\
\mathrm{QFI}_{yy} & = 4 \left( \langle \hat g_y^2 \rangle - \langle \hat g_y \rangle^2 \right)  
          = \frac{4k^2}{z_0^2} \left( \langle v^2 \rangle - \langle v \rangle^2 \right) \, , \\
\mathrm{QFI}_{zz} & = 4 \left( \langle \hat g_z^2 \rangle - \langle \hat g_z \rangle^2 \right)   
          = \frac{k^2}{z_0^4} \left( \langle ( u^2 + v^2 )^2 \rangle - \langle u^2 + v^2 \rangle^2 \right) \, , \\
\mathrm{QFI}_{xy} = \mathrm{QFI}_{21} & = 4 \left( \langle \hat g_x \hat g_y \rangle - \langle \hat g_x \rangle \langle \hat g_y \rangle \right)   
          = \frac{4k^2}{z_0^2} \left( \langle u v \rangle - \langle u \rangle \langle v \rangle \right) \, , \\
\mathrm{QFI}_{xz} = \mathrm{QFI}_{31} & = 4 \left( \langle \hat g_x \hat g_z \rangle - \langle \hat g_x \rangle \langle \hat g_z \rangle \right)
          = \frac{2k^2}{z_0^3} \left( \langle u ( u^2 + v^2 ) \rangle - \langle u \rangle \langle u^2 + v^2 \rangle \right) \, , \\
\mathrm{QFI}_{yz} = \mathrm{QFI}_{32} & = 4 \left( \langle \hat g_y \hat g_z \rangle - \langle \hat g_y \rangle \langle \hat g_z \rangle \right)
          = \frac{2k^2}{z_0^3} \left( \langle v ( u^2 + v^2 ) \rangle - \langle v \rangle \langle u^2 + v^2 \rangle \right) \, . \label{QFIm6}
\end{align}

Here we have used the notation:
\begin{align}
\langle f(u,v) \rangle := \langle \psi(0) | f(\hat u, \hat v) | \psi(0)\rangle 
= \frac{1}{N_C} \sum_u f(u, v) \, .  
\end{align}

\subsection{Two sources}

Consider a system of two point-like emitters.
In this case $M$ is a $ 2 \times 2$ matrix. 
We have
\begin{align}
M = C(r)^\dag C(r')
\end{align}
with
\begin{align}
C(r) = \left(
\begin{array}{cc}
c(w_1,r_1)   & c(w_1,r_2) \\
c(w_2,r_1)   & c(w_2,r_2) \\
\dots      & \dots    \\
c(w_{N_C},r_1) & c(w_{N_C},r_2)  
\end{array}
\right) \, ,
\end{align}
and
\begin{align}
C(r') = \left(
\begin{array}{cc}
c(w_1,r'_1)   & c(w_1,r'_2) \\
c(w_2,r'_1)   & c(w_2,r'_2) \\
\dots      & \dots    \\
c(w_{N_C},r'_1) & c(w_{N_C},r'_2)  
\end{array}
\right) \, .
\end{align}

Using the Dirac notation, we identify each column as a vector, i.e.,
\begin{align}
C(r) & = \left(
\begin{array}{cc}
|\psi(r_1)\rangle & |\psi(r_2)\rangle 
\end{array}
\right) \, , \\
C(r') & = \left(
\begin{array}{cc}
|\psi(r'_1)\rangle & |\psi(r'_2)\rangle
\end{array}
\right) \, ,
\end{align}
and the matrix $M$ then reads
\begin{align}
M = \left(
\begin{array}{cc}
\langle \psi(r_1) | \psi(r'_1) \rangle & \langle \psi(r_1) | \psi(r_2) \rangle \\
\langle \psi(r_2) | \psi(r'_1) \rangle & \langle \psi(r_2) | \psi(r'_2) \rangle 
\end{array}
\right) \, .
\end{align}

\subsubsection{Estimating the separation between the sources}

Consider a symmetric setup in which the two sources have 
coordinates $r_1 = - r_2$. Our goal is to estimate the separation $\Delta x$ along
one given direction, for example along the coordinate $x$.
Below $\delta\Delta x$ denotes a small variation of this parameter.

We have:
\begin{align}
|\psi(r'_1)\rangle & = e^{-i \hat g_x \frac{\delta\Delta x}{2} } |\psi(r_1)\rangle \, , \\
|\psi(r'_2)\rangle & = e^{i \hat g_x \frac{\delta\Delta x}{2} } |\psi(r_2)\rangle \, .
\end{align}

This implies
\begin{align}
\langle \psi(r_1) |\psi(r'_1) \rangle 
= \langle \psi(r_1) | e^{-i \hat g_x \frac{\delta\Delta x}{2} } |\psi(r_1)\rangle 
=  
\sum_j e^{-i k \frac{u_j}{z_0} \frac{\delta\Delta x}{2} } \, ,
\end{align}
and
\begin{align}
\langle \psi(r_2) |\psi(r'_2)\rangle = \langle \psi(r_2) | e^{i \hat g_x \frac{\delta\Delta x}{2}} |\psi(r_2)\rangle 
=  
\sum_j e^{i k \frac{u_j}{z_0} \frac{\delta\Delta x}{2}} \, .
\end{align}
Therefore
\begin{align}
\langle \psi(r_2) |\psi(r'_2)\rangle = \langle \psi(r_1) |\psi(r'_1)\rangle^*  \, .
\end{align}

Also note that
\begin{align}
\langle \psi(r_1)| \psi(r'_2) \rangle & = \langle \psi(r_1)| e^{i \hat g_x \frac{\delta\Delta x}{2}} |\psi(r_2) \rangle \\
& = \langle \psi(r_2)| e^{-i \hat g_x \frac{\delta\Delta x}{2}} |\psi(r_1) \rangle^* 
= \langle \psi(r_2)| \psi(r'_1) \rangle^* \, .
\end{align}

The above implies that the matrix $M$ has the general form:
\begin{align}
M = \left(
\begin{array}{cc}
\alpha & \beta \\
\beta^* & \alpha^*
\end{array}
\right) \, .
\end{align}
We can readily compute the trace norm for a matrix of this form (under the condition
$|\alpha| > |\beta|$):
\begin{align}
\| M \|_1 = 2 |\alpha| \, .
\end{align}

Since $\alpha = \sum_j e^{-i k \frac{u_j}{z_0} \frac{\delta\Delta x}{2}}$,
we have
\begin{align}
\| M \|_1 =
2 \left| \sum_j e^{-i k \frac{u_j}{z_0} \frac{\delta\Delta x}{2}} \right| \, .
\end{align}

It remains to expand this quantity up to the second-order in $\delta\Delta x$:
\begin{align}
\| M \|_1 & \simeq 2 \left| N_C - i k \frac{\delta\Delta x}{2z_0} \sum_j u_j 
- \frac{k^2}{2}\frac{\delta\Delta x^2}{4z_0^2} \sum_j u_j^2 \right| \\
& \simeq 2 N_C \left[ 1 - \frac{k^2}{2}\frac{\delta\Delta x^2}{4z_0^2} \frac{1}{N_C} \sum_j u_j^2 
+ \frac{k^2}{2} \frac{\delta\Delta x^2}{4z_0^2} \left(  \frac{1}{N_C} \sum_j u_j \right)^2 \right] \, ,
\end{align}
and finally
\begin{align}
\frac{1}{2 N_C} \| M \|_1 & = 
1 - \frac{k^2}{8} \frac{\delta\Delta x^2}{z_0^2} \left[
\frac{1}{N_C} \sum_j u_j^2 - \left( \frac{1}{N_C}\sum_j u_j \right)^2 \right] \\
& = 
1 - \frac{k^2}{8} \frac{\delta\Delta x^2}{z_0^2} \left[
\langle u^2 \rangle - \langle u \rangle^2 \right] \, .
\end{align}

From this we directly obtain the quantum Fisher information for the estimate of $\Delta x_1$:
\begin{align}
I_Q(\Delta x) & = \frac{8\left( 1 - \frac{1}{2N_C}\| M \|_1 \right)}{\delta\Delta x^2} 
= \frac{k^2}{z_0^2} \left( \langle u^2 \rangle - \langle u \rangle^2 \right) \, .
\end{align}

We can similarly obtain the quantum Fisher information for the separation in
any direction. It follows that also, in this case, the quantum Fisher information
matrix is proportional to the covariance matrix of the infinitesimal generators.
The elements of the QFI matrix are therefore equal, up to a multiplicative factor $4$, to the QFI matrix elements for the coordinates of a single emitter in Eqs.\ (\ref{QFIm1})-(\ref{QFIm6}).

\subsubsection{Transverse localization of the centroid}

We assume a configuration of the collectors that is inversion-symmetric.
This means that for each collector at location $w' \neq (0,0)$ there is another collector
at location $-w'$. This in turn implies that the amplitude
\begin{align}
\langle \psi(0) | U(r) | \psi(0) \rangle = 
\sum_u e^{-i k \frac{u x + v y}{z_0} }
\end{align}
is real for any $U(r)$.
Explicitly we have:
\begin{align}
\langle \psi(0) | U(r) | \psi(0) \rangle =
\left\{
\begin{array}{cl}
2 \sum_{j|w_j \neq (0,0)} \cos{ \left( k \frac{u_j x + v_j y}{z_0} \right) } & \mbox{if} \, \, N_C \, \, \mbox{is even} \, , \\
1 + 2 \sum_{j|w \neq (0,0)} \cos{ \left( k \frac{u_j x + v_j y}{z_0} \right) } & \mbox{if} \, \, N_C \, \, \mbox{is odd} \, .
\end{array} 
\right.
\end{align}

Consider an infinitesimal transverse displacement,
for example along the first coordinate direction $x$.
We have
\begin{align}
\langle \psi(0) | U(r) e^{- i \hat g_x \delta x } | \psi(0) \rangle 
& \simeq  \langle \psi(0) | U(r) \left( 1 - i \hat g_x \delta x - \hat g_x^2 \frac{\delta x^2}{2} \right) | \psi(0) \rangle \\
& \simeq  \langle \psi(0) | U(r) | \psi(0) \rangle 
+ \delta x \frac{\partial}{\partial x} \langle \psi(0) | U(r) | \psi(0) \rangle 
+ \frac{\delta x^2}{2} \frac{\partial^2}{\partial x^2} \langle \psi(0) | U(r) | \psi(0) \rangle\\
& = 
p + 2 \sum_{j|w \neq (0,0)} \cos{ \left( k \frac{u_j x + v_j y}{z_0} \right) }
- 2 k \frac{\delta x}{z_0} \sum_{j|w_j \neq (0,0)} u_j \sin{ \left( k \frac{u_j x + v_j y}{z_0} \right) } \nonumber \\
& - k^2 \frac{\delta x^2}{z_0^2} \sum_{j|w_j \neq (0,0)} u_j^2 \cos{ \left( k \frac{u_j x + v_j y}{z_0} \right) } \, ,
\end{align}
where $p=1$ is $N_C$ is odd, and $p=0$ otherwise.

We will write this second order expansion of the amplitude as
\begin{align}
\langle \psi(0) | U(r) e^{- i \hat g_x \delta x } | \psi(0) \rangle 
& \simeq  
A(r) - B(r) \delta x - C(r) \delta x^2 \, .
\end{align}

Consider now the matrix $M$ for a transverse displacement of both sources. We denote as $r_1$ the vector of coordinates of the first source, and 
as $r_2$ the coordinates of the second. The relative coordinate is
$\Delta r = r_1 - r_2$.
The matrix $M$ then reads as (up to the second order in $\delta x$)
\begin{align}
M = \left(
\begin{array}{cc}
N_C - C(0) \delta x^2                                            & A(\Delta r) - B(\Delta r) \delta x - C(\Delta r) \delta x^2 \\
A(\Delta r) + B(\Delta r) \delta x - C(\Delta r) \delta x^2  & N_C - C(0) \delta x^2
\end{array}
\right) \, .
\end{align}

For this matrix we can compute the trace norm directly:
\begin{align}
\frac{1}{2N_C} \| M \|_1 & = 
1 - \frac{1}{N_C} C(0) \delta x^2 + \frac{1}{2 N_C^2} B(\Delta r)^2 \delta x^2 \\
& = 1 - \frac{k^2 \delta x^2 }{z_0^2} \left[ \frac{1}{N_C} \sum_{j|w_j \neq (0,0)} {u_j}^2 \right] 
      - \frac{1}{2 N_C^2} \frac{k^2 \delta x^2 }{z_0^2} \left[ 2 \sum_{j|w_j \neq (0,0)} u_j \sin{\left( k \frac{u_j \Delta x + v_j \Delta y}{z_0} \right)} \right]^2 \\
& = 1 - \frac{k^2 \delta x^2 }{z_0^2} \left[ \frac{1}{N_C} \sum_{j} {u_j}^2 \right] 
      - \frac{1}{2 N_C^2} \frac{k^2 \delta x^2 }{z_0^2} \left[ 2 \sum_{j} u_j \sin{\left( k \frac{u_j \Delta x + v_j \Delta y}{z_0} \right)} \right]^2 \, .
\end{align}
Note that the last term is proportional to $\Delta r^2$, therefore it can be neglected in the paraxial approximation. We then have
\begin{align}
\frac{1}{2N_C} \| M \|_1 & \simeq 
1 - \frac{k^2 \delta x^2}{z_0^2} \left[ \frac{1}{N_C} \sum_{j} {u_j}^2 \right] \, .
\end{align}

Finally, we obtain an expression for the quantum Fisher information for the transverse coordinate of the centroid:
\begin{align}
I_Q(x) & = \lim_{\delta x \to 0} \frac{8 \left( 1 - \frac{1}{2 N_C} \| M \|_1 \right)}{\delta x^2} \\
    & = \frac{4k^2}{z_0^2} \left( \frac{1}{N_C} \sum_{j} {u_j}^2 \right) 
= \frac{4k^2}{z_0^2} \langle u^2 \rangle \, .
\end{align}

In conclusion, also, in this case, we obtain that the QFI for the coordinate of the centroid is proportional to the variance of the corresponding generators.

The analogous result is obtained for the coordinate of the centroid along any transverse direction. This implies that the QFI matrix for the transverse estimation of the centroid equals the corresponding sub-matrix of the single-emitter QFI matrix, whose elements are shown in Eqs.\ (\ref{QFIm1})-(\ref{QFIm6}).


\section{Physical implementation}
\label{sec:phys_imp}

\subsection{Two detectors}
\label{sec:two_det}

Consider a schematic of the two sources and of the measurement scheme as depicted in Fig.~\ref{f:schematic}. The two sources have coordinates $(\pm {\Delta x}/{2}, 0, {\pm  z}/{2} + z_0)$

\begin{figure}[h!]\center
\includegraphics[trim = 0cm 0cm 0cm 0cm, clip, width=0.5\linewidth]{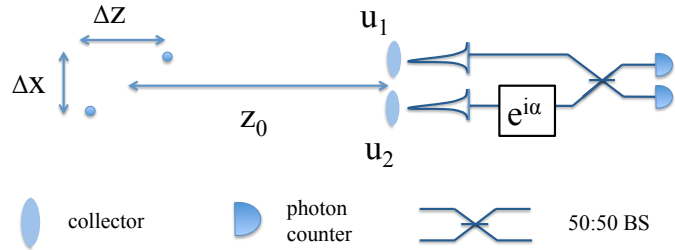} 
  \caption{Schematic of the two sources, they have a separation of $\Delta x$ on the object plane, and the object planes are separated by a distance $\Delta z$ in the axial direction to the collectors. The collectors are separated by a distance $|u_1-u_2|$.  }\label{f:schematic}
\end{figure}

Making the approximation that $\Delta x , \Delta z, |u_1|, |u_2| \ll z_0$, the QFIs for estimating the separation are
\begin{align}
\mathrm{QFI}_{\Delta x}  & \approx k^2\frac{(u_1- u_2)^2}{4 z_0^2}.    \label{eq:QFI_bucketl} \\
\mathrm{QFI}_{{\Delta z}}  & \approx k^2\frac{(u_1-u_2)^2 \left(\Delta x^2+ (u_1+u_2)^2\right)}{16 z_0^4}
   \label{eq:QFI_buckets}
\end{align}

The optimal measurement, also shown in Fig.~ \ref{f:schematic}, is a simple text-book example of interferometry.

Assuming the operators transform as
\begin{align}
    a^{\dagger'} _{u_1} \rightarrow\frac{1}{\sqrt2}(a^\dagger_{u_1} + a^\dagger_{u_2}) \nn
        a^{\dagger'}_{u_2} \rightarrow\frac{1}{\sqrt2}(a^\dagger_{u_1} - a^\dagger_{u_2}),
\end{align}
\noindent the probabilities of measuring a photon at modes 1 and 2 are respectively
\begin{align}
p_1= \frac{1}{4} \bigg[2+\cos \left(\frac{(u_1-u_2) (\Delta z (\Delta x - u_1 - u_2) + 2 \Delta x z_0)}{4 z_0^2} + \alpha\right) \nn
+\cos \left(\frac{(u_1-u_2) (\Delta z (\Delta x + u_1 + u_2) - 2 \Delta x z_0)}{4 z_0^2}+\alpha\right)\bigg] \nn
p_2 =\frac{1}{4} \bigg[2-\cos \left(\frac{(u_1-u_2) (\Delta z (\Delta x - u_1 - u_2)+2 \Delta x z_0)}{4 z_0^2}+\alpha\right) \nn
-\cos \left(\frac{(u_1-u_2) (\Delta z (\Delta x + u_1 + u_2) - 2 \Delta x z_0)}{4 z_0^2}+\alpha\right)\bigg.
\label{eq:2_det_ps}
\end{align}

If $\theta$ the unknown parameter to be estimated, given a set of measurement outcomes $\{x\}$, each occuring with probability $p(x|\theta)$, the CFI of $\theta$ is
\begin{align}
F_\theta = \sum_x p(x|\theta) \left[ \frac{\partial \log{p(x|\theta)}}{\partial \theta} \right]^2 \, ,
\label{eq:fi}
\end{align}


For the parameter $\Delta x$, applying a relative phase shift $\alpha = \theta_s$, substituting Eq~\eqref{eq:2_det_ps} into Eq~\eqref{eq:fi}, the CFI is 

\begin{align}
F_{\Delta x}= k^2
\frac{\left(\frac{(u_1-u_2) \sin \left(\frac{\theta_s \Delta z+\Delta x u_1-\Delta x u_2+u_1^2-u_2^2+2 \theta_s z_0}{\Delta z+2 z_0}\right)}{\Delta z+2 z_0}+\frac{(u_1-u_2) \sin \left(\frac{\theta_s \Delta z+\Delta x u_1-\Delta x u_2-u_1^2+u_2^2-2 \theta_s z_0}{\Delta z-2 z_0}\right)}{\Delta z-2 z_0}\right)^2}
{4 \left(-\cos \left(\frac{\theta_s \Delta z+s u_1-\Delta x u_2+u_1^2-u_2^2+2 \theta_s z}{\Delta z+2 z_0}\right)-\cos \left(\frac{\theta_s \Delta z+\Delta x u_1-s u_2-u_1^2+u_2^2-2 \theta_s z}{\Delta z-2 z_0}\right)+2\right)}
\\
+k^2\frac{\left(-\frac{(u_1-u_2) \sin \left(\frac{\theta_s +\Delta x u_1-\Delta x u_2+u_1^2-u_2^2+2 \theta_s z}{\Delta z+2 z_0}\right)}{\Delta z+2 z_0}-\frac{(u_1-u_2) \sin \left(\frac{\theta_s \Delta z+\Delta x u_1-\Delta x u_2-u_1^2+u_2^2-2 \theta_s z}{\Delta z-2 z_0}\right)}{\Delta z-2 z_0}\right)^2}{4 \left(\cos \left(\frac{\theta_s \Delta z+\Delta x u_1-\Delta x u_2+u_1^2-u_2^2+2 \theta_s z}{\Delta z+2 z_0}\right)+\cos \left(\frac{\theta_s \Delta z+\Delta x u_1-\Delta x u_2-u_1^2+u_2^2-2 \theta_s z}{\Delta z-2 z_0}\right)+2\right)}.
\end{align}
\noindent Given $\Delta z,  \ll z$ and 
$[\Delta x (u_1 - u_2)]/(2 z_0) \ll 1$, setting $\theta_s = 0$ gives  
\begin{align}
F_{\Delta x}\approx \frac{(u_1-u_2)^2}{(4z_0^2)},
\end{align}
\noindent which is equal to the QFI. It is worth noting that our expressions for the QFI and CFI are for estimating the actual value of $\Delta x$, instead of the angular separation, which would be 
${\Delta x}/{z_0}$. If we were to estimate the angular separation, the expression for the QFI and CFI will increase by a factor of $z_0^2$, i.e
\begin{align}
    \text{QFI}_{\Delta x/z_0}\approx F_{\Delta x/z_0} \approx  \frac{(u_1-u_2)^2}{4}.
\end{align}
\noindent We thus reproduce the result of Tsang, Nair and Lu, namely that the QFI is constant and only depends on the properties of the imaging system.

For the parameter $\Delta z$, applying a relative phase shift $\alpha =\theta_\ell$, CFI is 
\begin{align}
F_{\Delta z} &= -k^2 (u_1-u_2)^2 \times\nn
&\bigg[\frac{(\Delta x+u_1+u_2) \sin \left(\frac{\theta_\ell \Delta z+\Delta x u_1-\Delta x u_2+u_1^2-u_2^2+2 \theta_\ell z}{\Delta z+2 z_0}\right)}{(\Delta z+2 z_0)^2}+ 
\frac{(\Delta x-u_1-u_2) \sin \left(\frac{\theta_\ell \Delta z+\Delta x u_1-\Delta x u_2-u_1^2+u_2^2-2 \theta_\ell z}{\Delta z-2 z_0}\right)}{(\Delta z-2 z_0)^2} \bigg]^2 \times \nn
& \bigg\{\left(\cos \left(\frac{\theta_\ell \Delta z+\Delta x u_1-\Delta x u_2+u_1^2-u_2^2+2 \theta_\ell z}{\Delta z+2 z_0}\right)+\cos \left(\frac{\theta_\ell \Delta z+\Delta x u_1-\Delta x u_2-u_1^2+u_2^2-2 \theta_\ell z}{\Delta z-2 z_0}\right)-2\right) 
 \nn
 & \left(\cos \left(\frac{\theta_\ell \Delta z+\Delta x u_1-\Delta x u_2+u_1^2-u_2^2+2 \theta_\ell z}{\Delta z+2 z_0}\right)+\cos \left(\frac{\theta_\ell \Delta z+\Delta x u_1-\Delta x u_2-u_1^2+u_2^2-2 \theta_\ell z}{\Delta z-2 z_0}\right)+2\right)\bigg\}^{-1}
\end{align}

In general, $\theta_s \neq \theta_l$, which is logical because here there are 2 parameters to extract, but there is only one variable in the measurement outcome.
The optimal phase for estimating $\Delta z$ at $\Delta z \approx 0$ is

\begin{align}
\theta_\ell=
\tan ^{-1}\bigg[&\frac{\Delta x \cos \left(\frac{\Delta x (u_1-u_2)}{2 z_0}\right)}{\sqrt{\Delta x^2 \cos ^2\left(\frac{\Delta x (u_1-u_2)}{2 z_0}\right)+(u_1+u_2)^2 \sin ^2\left(\frac{\Delta x (u_1-u_2)}{2 z_0}\right)}},\nn
 -&\frac{\sqrt{2} (u_1+u_2) \sin \left(\frac{\Delta x (u_1-u_2)}{2 z_0}\right)}{\sqrt{\left(\Delta x^2-(u_1+u_2)^2\right) \cos \left(\frac{\Delta x (u_1-u_2)}{z_0}\right)+\Delta x^2+(u_1+u_2)^2}}\bigg].
\end{align}

\subsection{Four collectors - simultaneous estimation of transverse and axial separations }

\begin{figure}[h!] \center
\includegraphics[trim = 0cm 0cm 0cm 0cm, clip, width=0.6\linewidth]{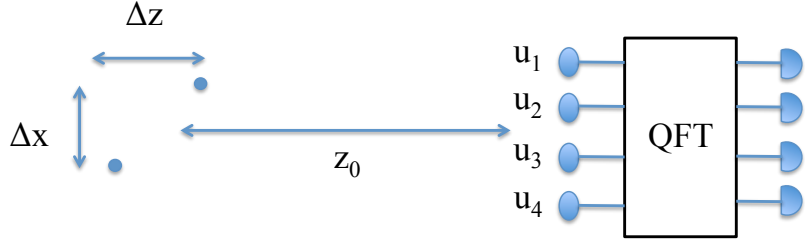} 
  \caption{\label{f:4_det} Schematic for estimating simultaneously $\Delta x$ and $\Delta z$ using 4 collectors which are evenly spaced, centered at position 0. The optimal linear optical transformation is a 4-mode quantum Fourier transform.}
\end{figure}

We have seen in Sec.~\ref{sec:two_det} that using only two collectors, one cannot simultaneously optimally measure the transverse and axial separations. Intuitively, this is due to the fact that there are two parameters to extract and only one variable that changes with the measurement.

Since we know the commutators for the different parameters to be estimated (using Eqs.~\eqref{eq:gens}), we can configure the positions of
the collectors such that the off-diagonal terms in the QFI matrix are zero. If such a condition is satisfied, then there exists a measurement that can simultaneously. One such configuration is depicted in Fig~\ref{f:4_det}, where the collectors are evenly spaced, with positions, where the center is at 0.

For the parameter $\Delta x$, the QFI for putting the four collectors along the $u_1$ axis in generic positions $(u_1,u_2,u_3,u_4)$ is
\begin{align}
\text{QFI}_{\Delta x}(u_1,u_2,u_3,u_4)=\frac{1}{16} \left(3 u_1^2-2 u_1 ({u_2}+{u_3}+{u_4})+3 {u_2}^2-2 {u_2} ({u_3}+{u_4})+3 {u_3}^2-2 {u_3} {u_4}+3 {u_4}^2\right).
\end{align}

Now setting them evenly spaced, the coordinates are $\left(u_1, \frac{1}{3}u_1, -\frac{1}{3}u_1, - u_1\right)$, the QFI is
\begin{align}
   \text{ QFI}_{\Delta x}\left(u_1, \frac{1}{3}u_1, -\frac{1}{3}u_1, - u_1\right) = 
   \frac{5}{9z_0^2}u_1^2.
\end{align}

\noindent
We then apply a 4-mode quantum Fourier transform, which
acts on the 4 modes as
\begin{align} 
\left(
\begin{array}{c}
 a^{\dagger'} _{u_1} \\
 a^{\dagger'} _{u_2} \\
 a^{\dagger'} _{u_3} \\
 a^{\dagger'} _{u_4} \\
\end{array}
\right) \rightarrow
    \frac{1}{2}\left(
\begin{array}{cccc}
 1 & 1 & 1 & 1 \\
 1 & i & -1 & -i \\
 1 & -1 & 1 & -1 \\
 1 & -i & -1 & i \\
\end{array}
\right)
\left(
\begin{array}{c}
 a^\dagger_{u_1} \\
 a^\dagger_{u_2} \\
 a^\dagger_{u_3} \\
 a^\dagger_{u_4} \\
\end{array}
\right).
\end{align}

\noindent The probability of the photon exiting through the mode 1 is
\begin{align}
p_1 = \frac{1}{16} 
\bigg[&\cos \left(\frac{(u_1-u_2) (\Delta z (-\Delta x+u_1+u_2)+2 \Delta x z_0)}{4 z_0^2}\right)
     +\cos \left(\frac{(u_1-u_2) (\Delta z (\Delta x+u_1+u_2)+2 \Delta x z_0)}{4 z_0^2}\right) \nn
     &+\cos \left(\frac{(u_1-u_3) (\Delta z (-\Delta x+u_1+u_3)+2 \Delta x z_0)}{4 z_0^2}\right)
     +\cos \left(\frac{(u_1-u_3) (\Delta z (\Delta x+u_1+u_3)+2 \Delta x z_0)}{4 z_0^2}\right) \nn
     &+\cos \left(\frac{(u_1-u_3) (\Delta z (-\Delta x+u_1+u_3)+2 \Delta x z_0)}{4 z_0^2}\right)
     +\cos \left(\frac{(u_1-u_3) (\Delta z (\Delta x+u_1+u_3)+2 \Delta x z_0)}{4 z_0^2}\right)\nn
     &+\cos \left(\frac{(u_2-u_3) (\Delta z (-\Delta x+u_2+u_3)+2 \Delta x z_0)}{4 z_0^2}\right)
     +\cos \left(\frac{(u_2-u_3) (\Delta z (\Delta x+u_2+u_3)+2 \Delta x z_0)}{4 z_0^2}\right)\nn
     &+\cos \left(\frac{(u_2-u_3) (\Delta z (-\Delta x+u_2+u_3)+2 \Delta x z_0)}{4 z_0^2}\right)
     +\cos \left(\frac{(u_2-u_3) (\Delta z (\Delta x+u_2+u_3)+2 \Delta x z_0)}{4 z_0^2}\right)\nn
     &+\cos \left(\frac{(u_3-u_3) (\Delta z (-\Delta x+u_3+u_3)+2 \Delta x z_0)}{4 z_0^2}\right)
     +\cos \left(\frac{(u_3-u_3) (\Delta z (\Delta x+u_3+u_3)+2 \Delta x z_0)}{4 z_0^2}\right) 
     +4\bigg]
\end{align}
\noindent The rest of the probabilities take a similar form, which we omit for brevity.
\noindent
Once again, calculating the probabilities and using Eq.~\eqref{eq:fi}, the Fisher information for photon counting at the output ports is
\begin{align}
\text{CFI}_{\Delta x} &= \frac{1}{18} u_1^2 \left[\cos \left(\frac{2 \Delta x u_1}{3}\right)+9\right] \\
     &\approx \frac{5}{9z_0^2}u_1^2
     \label{eq:u1},
\end{align}

\noindent with Eq~\eqref{eq:u1} coinciding with the QFI.

\noindent
In the same configuration, the QFI for $\Delta z$ is
\begin{align}
\text{QFI}_{\Delta z}=\frac{1}{z_0^4}\left(\frac{5 \Delta x^2 u_1^2}{36}+\frac{4 u_1^4}{81} \right).
\end{align}

\noindent
The actual expression for the CFI of $\Delta z$ is large and un-illuminating. However, in the limit that 
$ \Delta x \rightarrow0$,
it reduces to
\begin{align}
\text{CFI}_{\Delta z} = \frac{4}{81 z_0^4} u_1^4
\end{align}
\noindent which coincides with the QFI. 
In this limit both the parameters $\Delta x$ and $\Delta z$ can be extracted here optimally simultaneously.

\end{widetext}

\end{document}